\documentclass[aps,prd,epsf,amsmath,amssymb,graphics,preprint]{revtex4}
\usepackage{graphicx}% Include figure files
\usepackage{dcolumn}% Align table columns on decimal point
\usepackage{bm}% bold math

\usepackage{color}

\newcommand{\Uout}{U_{\rm out}}
\newcommand{\Vin}{V_{\rm in}}
\newcommand{\uout}{u_{\rm out}}
\newcommand{\vin}{v_{\rm in}}

\newcommand{\dUout}{\dot{U}_{\rm out}}
\newcommand{\dVin}{\dot{V}_{\rm in}}
\newcommand{\duout}{\dot{u}_{\rm out}}
\newcommand{\dvin}{\dot{v}_{\rm in}}

\newcommand{\tauo}{\tau_{\rm out}}
\newcommand{\taui}{\tau_{\rm in}}

\newcommand{\etaend}{\eta_{\rm e}}
\newcommand{\etabgn}{\eta_{\rm b}}

\newcommand{\etar}{\eta_{\rm rf}}

\newcommand{\Us}{U_{\rm s}}
\newcommand{\Vs}{V_{\rm s}}
\newcommand{\us}{u_{\rm s}}
\newcommand{\vs}{v_{\rm s}}

\newcommand{\dUs}{\dot{U}_{\rm s}}
\newcommand{\dVs}{\dot{V}_{\rm s}}
\newcommand{\dus}{\dot{u}_{\rm s}}
\newcommand{\dvs}{\dot{v}_{\rm s}}

\newcommand{\ts}{t_{\rm s}}
\newcommand{\rs}{r_{\rm s}}
\newcommand{\etas}{\eta_{\rm s}}
\newcommand{\chis}{\chi_{\rm s}}

\newcommand{\rstar}{r_{*}}

\newcommand{\as}{a_{\rm s}}

\newcommand{\etab}{\eta_{\rm S}}
\newcommand{\etaf}{\eta_{\rm G}}

\newcommand{\drs}{\dot{r}_{\rm s}}
\newcommand{\dchis}{\dot{\chi}_{\rm s}}

\newcommand{\chii}{\chi_{\rm i}}

\newcommand{\rf}{R_{\rm G}}
\newcommand{\rb}{R_{\rm S}}

\newcommand{\etac}{\eta_{\rm c}}
\newcommand{\tc}{t_{\rm c}}

\newcommand{\Aout}{A_{\rm out}}
\newcommand{\Bin}{B_{\rm in}}
\newcommand{\Cout}{C_{\rm out}}
\newcommand{\Din}{D_{\rm in}}

\begin{document}

\title{Radiative gravastar with Gibbons-Hawking temperature}

\author{Ken-ichi Nakao$^{1,2}$}
\author{Kazumasa Okabayashi$^1$}
\author{Tomohiro Harada$^3$}

\affiliation{
${}^{1}$
Department of Mathematics and Physics, Graduate School of Science, Osaka City University, 3-3-138 Sugimoto, Sumiyoshi, Osaka 558-8585, Japan\\
${}^{2}$Nambu Yoichiro Institute of Theoretical and Experimental Physics,
Osaka City University, Sumiyoshi, Osaka City 558-8585, Japann\\
${}^{3}$Department of Physics, Rikkyo University, Toshima, Tokyo 171-8501, Japan}

%%%%%%%%%%%%%%%%%%%%%%%%%%%%%%%%%%%%%%%%%
%%%%%%%%%%%%%%%%%%%%%%%%%%%%%%%%%%%%%%%%%

\begin{abstract}
We study the quantum particle creation in a toy model of spherically symmetric gravitational collapse whose final product is not 
a black hole but a gravastar. 
Precedent studies revealed that even in the case of the gravitational collapse to form a horizonless ultra-compact object, 
thermal radiation named transient Hawking radiation is generated at the late stage of the gravitational collapse, and   
a sudden stop of collapsing motion to form a horizonless ultra-compact object causes one or two bursts of quantum particle creation. 
The very different behavior of the model studied in this paper from the precedent ones is quantum radiation 
with a thermal spectrum from the gravastar between two bursts.  
The temperature of the radiation is not the same as the Hawking one determined by the gravitational mass of the system 
but the Gibbons-Hawking one of the de Sitter core inside the gravastar.

\end{abstract}

\preprint{OCU-PHYS-560}
\preprint{AP-GR-180}
\preprint{NITEP 133}
\preprint{RUP-22-7}
\date{\today}
\maketitle

%\newpage
%%%%%%%%%%%%%%%%%%%%%%%%%%%%%
%%%%%%%%%%%%%%%%%%%%%%%%%%%%%
%%%%%%%%%%%%%%%%%%%%%%%%%%%%%
\section{Introduction}
%%%%%%%%%%%%%%%%%%%%%%%%%%%%%
%%%%%%%%%%%%%%%%%%%%%%%%%%%%%
%%%%%%%%%%%%%%%%%%%%%%%%%%%%%

The gravitational collapse to form black holes is one of the most important and fascinating subjects in the physics of strong gravity 
and has gathered special interests. 
Recent observational developments have led to a more activity research field of the black hole physics; 
the observations through gravitational 
waves \cite{LIGO:2016,LVK:2021} and the imaging of the so-called black hole shadow \cite{EHT:2019} gave us very impressive evidence 
of the existence of black hole candidates in our universe. 

The black hole is defined as a complement of the causal past of the future conformal infinity, and 
its boundary is called the event horizon \cite{Penrose:1969, Hawking:1970}. 
This definition captures the essence of the black hole and the event horizon. 
On the other hand, any observational confirmation of the black hole defined in such a manner is impossible \cite{Cardoso-P,Nakao-YH}.   
Exactly speaking, the gravitational waves observed by LIGO and Virgo were generated by not black hole binaries but 
collapsing objects; what was taken through EHT is not a direct image of a black hole but an image of 
collapsing matter (see, for example, Appendix A of Ref.~\cite{OAN}). 
This is true also for semi-classical processes which are quantum phenomena in classical spacetime. 
The Hawking radiation is a celebrated example which is regarded as a characteristic of the black hole \cite{Hawking:1974,Hawking:1975}. However,   
Paranjape and Padmanabhan studied a free quantum field in the gravitational collapse to form a static horizonless 
spherical ultra-compact object with a hollow inside and showed that even if no black hole eventually forms,  
radiation power whose value is the same as that of the Hawking radiation 
is generated at the late stage of the gravitational collapse \cite{PP:2009}.  
The study by Barcel\'{o}, Liberati, Sonego and Visser \cite{BLSV:2011} revealed that the 
quantum radiation studied by Paranjape and Padmanabhan has the thermal spectrum of the Hawking temperature. 
Even quantum radiation with thermal spectrum is not an evidence 
of a black hole but merely implies it is a black hole candidate.

What we can observationally confirm is that 
the black hole candidate is a black hole mimicker. The black hole mimicker is a horizonless ultra-compact object 
observationally very similar to a black hole. Various black hole mimickers have been proposed (see eg. Ref.~\cite{Cardoso-P}). 
The situation studied by Paranjape and Padmanabhan is regarded as a formation process of a black hole mimicker. 
One of the present authors, TH, in collaboration with Cardoso and Miyata, studied the quantum particle creation 
in the similar situation to that studied by Paranjape and Padmanabhan \cite{Harada-CM}. 
They showed that the two bursts of the particle creation occur due to the stop 
of the gravitational collapse in addition to the thermal radiation revealed 
by Paranjape and Padmanabhan. 
Kokubu and TH studied  the quantum field which interacts with the collapsing object \cite{Kokubu-H} 
in the spacetime similar to that studied in Ref.~\cite{Harada-CM}. 
They also studied the case in which the interaction between the quantum field and the collapsing object is represented 
as the reflection boundary condition for the mode function at the surface of the collapsing object.  
Their study revealed that, even in this case, 
the transient Hawking radiation is generated, and furthermore only one burst of the particle creation due to the stop of the 
collapse occurs. In Refs.~\cite{PP:2009,Harada-CM,Kokubu-H}, the gravitational collapse of 
the spherically symmetric object with hollow inside, or in other words, an infinitesimally thin spherical 
shell were studied. By contrast, the present authors studied the free quantum field in the gravitational 
collapse of the spherically symmetric object composed of homogeneously distributed 
matter enclosed by an infinitesimally thin crust, which will be called OHN model \cite{Okabayashi-HN}. 
The result obtained in Ref.~\cite{Okabayashi-HN} is basically the same as the results in Refs.~\cite{Harada-CM}: 
the homogeneous distribution of matter does not affect the time variation of the radiation power of the 
quantum particle creation even though the homogeneous component dominates the total mass of the system.  

The final product of OHN model is the spherically symmetric ultra-compact horizonless object composed of homogeneously 
distributed matter with infinitesimally thin crust. The equation of state of the homogeneous matter is $p=-\rho/3$, where $p$ and $\rho$ 
are the pressure and the energy density, respectively. By contrast, in this paper, we consider an OHM type model, but the final product 
is a gravastar. The gravastar is one of black hole mimickers which was proposed by Mazur and Mottola in order to solve problems related to 
the black hole (e.g., the information loss problem)\cite{MM2004}. Its inside is occupied by the dark energy of $p=-\rho<0$, or equivalently, 
the cosmological constant and hence the geometry of its inside is equivalent to that of the de Sitter spacetime.

This paper is organized as follows. In Sec.~II. we briefly review the quantum particle creation of 
the massless scalar field in the spherically symmetric spacetime and give a formula to calculate 
the radiation power in the spacetime with a homogeneous star. In Sec.~III, we explain the model of 
the gravastar formation through the gravitational collapse of a spherically symmetric dust star.  
In Sec.~IV, we show an example of quantum particle creation in the gravastar formation process, 
and briefly see what happens in the present model. In Sec.~V,  we estimate the 
radiation power in the radiative gravastar phase and its duration. Sec. VI is devoted to summary. 
Five Appendices are given at the end of the paper; Appendices A  and B are for the readers unfamiliar to this topics, whereas Appendices C, D and E 
show the detail of the calculations. 

We adopt the natural unit $c=\hbar=1$.  Newton's gravitational constant 
and the Boltzmann constant are denoted by $G$ and $k_{\rm B}$, respectively. The sign convention of the metric 
follows the text book written by Wald \cite{Wald}.

%%%%%%%%%%%%%%%%%%%%%%%%%%%%%
%%%%%%%%%%%%%%%%%%%%%%%%%%%%%
%%%%%%%%%%%%%%%%%%%%%%%%%%%%%
\section{Quantum particle creation}
%%%%%%%%%%%%%%%%%%%%%%%%%%%%%
%%%%%%%%%%%%%%%%%%%%%%%%%%%%%
%%%%%%%%%%%%%%%%%%%%%%%%%%%%%

We study the quantum dynamics of the massless scalar field $\phi$ in the spherically symmetric spacetime 
by the so called semi-classical treatment in which the effect of the quantum field on the classical 
spacetime geometry is ignored. 

%%%%%%%%%%%%%%%%%
%%%%%%%%%%%%%%%%%
\subsection{Radiation power}
%%%%%%%%%%%%%%%%%
%%%%%%%%%%%%%%%%%

We consider the quantum particle creation in the spherically symmetric asymptotically flat spacetime.  
By adopting the double null coordinates, the infinitesimal world interval can be written in the following form: 
\begin{equation}
ds^2=-h^2(u,v)dudv+r^2(u,v)d\Omega^2, \label{metric-0}
\end{equation}
where $u$ and $v$ are the retarded and the advanced time coordinates, respectively, whereas $d\Omega^2$ 
is the round metric. In the case of the Minkowski spacetime, $h=1$ and the areal radius $r$ 
agrees with $(v-u)/2$, and hence $u=v$ at the symmetry center $r=0$. 
By contrast, in general dynamical cases, the symmetry center $r=0$ is not $v=u$ but 
\begin{equation}
v=F(u). \label{sym-center}
\end{equation} 

In order to study the quantum effect in this spacetime, we consider the massless free scalar field which is the simplest model but is 
sufficient for the present purpose. The Lagrangian density is given as
\begin{equation}
{\cal L}=-\frac{1}{2}\sqrt{-g}g^{\mu\nu}(\partial_\mu\phi)\partial_\nu\phi,
\end{equation}
where $g_{\mu\nu}$, $g^{\mu\nu}$ and $g$ are the metric tensor, its inverse and its determinant, respectively. 
The average of radiation power of the massless scalar field due to the quantum effect caused by the spacetime curvature is estimated 
through the expectation value of the stress-energy-momentum tensor $T_{\mu\nu}$ as
\begin{align}
P(u)=\oint_{r\rightarrow\infty} \langle0|(\hat{T}_{uu}-\hat{T}_{vv})|0\rangle r^2d\Omega = \frac{1}{48\pi}\kappa^2(u), \label{P-def}
\end{align}
where, being a prime to be a derivative with respect to $u$, 
\begin{align}
\kappa(u):=-\frac{F''(u)}{F'(u)}, \label{kappa-def}
\end{align}
and we have ignored the total derivative term in $P$, since it does not contribute the total radiated energy. 
This expression is derived by invoking the S-wave approximation whose sketch is given in Appendix A. 
It is known that if the adiabatic condition $|\kappa'|/\kappa^2\ll1$ is satisfied, 
the spectrum of the radiation is thermal with the temperature \cite{BLSV:2011}
\begin{equation}
k_{\rm B}T=\frac{\kappa}{2\pi}.
\end{equation}

%%%%%%%%%%%%%%%%%
%%%%%%%%%%%%%%%%%
\subsection{How to calculate radiation power in the case of a homogeneous star}
%%%%%%%%%%%%%%%%%
%%%%%%%%%%%%%%%%%

The outside of the star is assumed to be vacuum. By Birkhoff's theorem, the outside domain is described 
by the Schwarzschild geometry whose metric is given as
\begin{align}
ds^2&=-f(r)dt^2+\frac{dr}{f(r)}+r^2 d\Omega^2 
=-f(r)\left(dt^2-dr_*^2\right)+r^2d\Omega^2,
\end{align}
where
\begin{equation}
f(r)=1-\frac{2GM}{r},
\end{equation}
and
\begin{equation}
r_*=\int \frac{dr}{f(r)}=r+2GM\ln\left|r-2GM\right|.
\end{equation}
The null coordinates are defined as
\begin{equation}
u=t-\rstar~~~~~{\rm and}~~~~~v=t+\rstar.
\end{equation}

As mentioned, the inside of the spherical star is assumed to be homogeneous and so described by 
the Robertson-Walker geometry whose metric is given as
\begin{align}
ds^2=a^2(\eta)\left[-d\eta^2+d\chi^2+\varSigma^2(\chi) d\Omega^2\right], \label{RW-metric}
\end{align}
where 
\begin{equation}
\varSigma(\chi)=
\begin{cases}
\sin\chi &\text{for positive curvature space,} \\
\chi      &\text{for flat space,} \\
\sinh\chi &\text{for negative curvature space.}
\end{cases}
\end{equation}
The null coordinates are defined as
\begin{equation}
U:=\eta-\chi~~~~{\rm and}~~~~V:=\eta+\chi.
\end{equation}

We denote the proper time along the world line of the surface of the star by $\tau$. 
Then by using the coordinates outside the star, the world line of the surface of the star, which is a curve with constant round coordinates, 
is represented in the form
\begin{align}
u&=\us(\tau)=\ts(\tau)-\rs(\tau)-2M\ln\left|\rs(\tau)-2GM\right|, \\
v&=\vs(\tau)=\ts(\tau)+\rs(\tau)+2M\ln\left|\rs(\tau)-2GM\right|,
\end{align}
whereas, by using the coordinates inside the star, the world line of the surface of the star is represented as 
\begin{align}
U&=\Us(\tau)=\etas(\tau)-\chis(\tau), \\
V&=\Vs(\tau)=\etas(\tau)+\chis(\tau).
\end{align}

Since the coordinate system we adopt here is different from that of Eq.~\eqref{metric-0}, we show how to derive $F''/F'$ in detail.  
The ingoing radial null with $v=\vin=$ constant and $V=\Vin=$ constant hits the surface of the star at $\tau=\taui$. 
After the radial null arrives at the origin $\chi=0$, 
it becomes the outgoing null with $U=\Uout=\Vin$ and $u=\uout$ and again hits the surface of the star at $\tau=\tauo$. Then, we denote
\begin{equation}
\dUout=\dot{\Us}(\tauo),~~\dVin=\dot{\Vs}(\taui),~~\duout=\dot{\us}(\tauo)~~{\rm and}~~\dvin=\dot{\vs}(\taui),
\end{equation}
where a dot represents a derivative with respect to the proper time $\tau$. Since we have
\begin{equation}
1=\frac{d\Vin}{d\Uout}=\frac{\dVin d\taui}{\dUout d\tauo},
\end{equation}
we obtain
\begin{equation}
\frac{\dVin}{\dUout}=\frac{d\tauo}{d\taui}.
\end{equation}
Then, we have
\begin{equation}
\frac{d\vin}{d\uout}=\frac{\dvin}{\duout}\frac{d\taui}{d\tauo}=\frac{\dvin}{\duout}\frac{\dUout}{\dVin}.
\end{equation}
We define $\Aout$ and $\Bin$ as
\begin{equation}
\Aout:=\frac{\dUout}{\duout}~~~~{\rm and}~~~~\Bin:=\frac{\dVin}{\dvin}.
\end{equation}
Then, we have
\begin{equation}
F'(\uout)=\frac{d\vin}{d\uout}=\frac{\Aout}{\Bin}.
\end{equation}
By differentiating this expression with respect to $u$, we have
\begin{align}
F''(\uout)&=\frac{dF'(\uout)}{d\uout}
=\frac{\Aout}{\Bin\duout}\left(\frac{d\ln\Aout}{d\tauo}-\frac{\dUout}{\dVin}\frac{d\ln\Bin}{d\taui}\right).
\end{align}
Hence, by defining $\Cout$ and $\Din$ as
\begin{align}
\Cout&:= -\frac{1}{\duout}\frac{d\ln\Aout}{d\tauo}, \\
\Din  &:=-\frac{1}{\dvin}\frac{d\ln\Bin}{d\taui},
\end{align}
we obtain
\begin{align}
\kappa(\uout)&:=-(\ln F')'(\uout)=\Cout-\frac{\Aout}{\Bin}\Din. \label{kappa-out}
\end{align}

The world line of the surface of the star is represented as
\begin{align}
\eta&=\etas(\tau), \\
\chi&=\chis(\tau)
\end{align}
in the inside coordinates, and
\begin{align}
t&=\ts(\tau), \\
r&=\rs(\tau),
\end{align}
in the outside coordinates. Denoting the scale factor at the surface $a(\etas)$ by $\as$, 
we have
\begin{align}
\frac{d\etas}{d\tau}&=\sqrt{\dchis^2+\frac{1}{\as^2}}, \\
\frac{d\ts}{d\tau}&=\frac{\sqrt{\drs^2+f(\rs)}}{f(\rs)},
\end{align}
and hence we obtain
\begin{align}
\dUs&=\sqrt{\dchis^2+\frac{1}{\as^2}}-\dchis, \\
\dVs&=\sqrt{\dchis^2+\frac{1}{\as^2}}+\dchis, \\
\dus&=\frac{\sqrt{\drs^2+f(\rs)}-\drs}{f(\rs)}, \\
\dvs&=\frac{\sqrt{\drs^2+f(\rs)}+\drs}{f(\rs)}.
\end{align}
We introduce the following quantities: 
\begin{align}
A&:=\frac{\dUs}{\dus}=\left(\sqrt{\drs^2+f(\rs)}+\drs\right)\left(\sqrt{\dchis^2+\frac{1}{\as^2}}-\dchis\right), \label{A-rep}\\
B&:=\frac{\dVs}{\dvs}=\left(\sqrt{\drs^2+f(\rs)}-\drs\right)\left(\sqrt{\dchis^2+\frac{1}{\as^2}}+\dchis\right), \\
C&:=-\frac{1}{\dus}\frac{d\ln A}{d\tau}=-\frac{1}{\dUs}\frac{dA}{d\tau}\nonumber \\
&=-\left(\frac{\ddot{r}_{\rm s}}{\sqrt{\drs^2+f(\rs)}}-\frac{\ddot{\chi}_{\rm s}}{\sqrt{\dchis^2+\dfrac{1}{\as^2}}}\right)
\left(\sqrt{\drs^2+f(\rs)}+\drs\right) \nonumber \\
&-\frac{GM\drs}{\rs^2\sqrt{\drs^2+f(\rs)}}
+\frac{1}{\as^3}\frac{d\as}{d\etas}\frac{\sqrt{\drs^2+f(\rs)}+\drs}{\sqrt{\dchis^2+\dfrac{1}{\as^2}}-\dchis}, \\
D&:=-\frac{1}{\dvs}\frac{d\ln B}{d\tau}=-\frac{1}{\dVs}\frac{dB}{d\tau}\nonumber \\
&=\left(\frac{\ddot{r}_{\rm s}}{\sqrt{\drs^2+f(\rs)}}-\frac{\ddot{\chi}_{\rm s}}{\sqrt{\dchis^2+\dfrac{1}{\as(\etas)}}}\right)
\left(\sqrt{\drs^2+f(\rs)}-\drs\right) \nonumber \\
&-\frac{GM\drs}{\rs^2\sqrt{\drs^2+f(\rs)}}
+\frac{1}{\as^3}\frac{d\as}{d\etas}\frac{\sqrt{\drs^2+f(\rs)}-\drs}{\sqrt{\dchis^2+\dfrac{1}{\as^2}}+\dchis}. 
\end{align}
Then, we have
\begin{equation}
\Aout=A(\tauo),~~\Bin=B(\taui),~~\Cout=C(\tauo),~~\Din=D(\taui). \label{ABCD-def}
\end{equation}

%%%%%%%%%%%%%%%%%%%%%%%%%%%%%
%%%%%%%%%%%%%%%%%%%%%%%%%%%%%
%%%%%%%%%%%%%%%%%%%%%%%%%%%%%
\section{Gravastar formation through dust collapse}
%%%%%%%%%%%%%%%%%%%%%%%%%%%%%
%%%%%%%%%%%%%%%%%%%%%%%%%%%%%
%%%%%%%%%%%%%%%%%%%%%%%%%%%%%

As already mentioned, we consider the gravastar formation through the gravitational collapse of a spherically symmetric star. 
The time evolution of the star is divided into three phases. In the first phase, the star is composed of homogeneously distributed 
dust and begins collapsing from the momentarily static configuration. We call this period {\it the collapsing phase}. 
In the second phase, the speed of the gravitational collapse slows down. This period is called {\it the slowing-down phase}. 
The ingredient of the star is not the  dust in this phase. 
Finally, the star stops collapsing and becomes a static gravastar. We call this period {\it the gravastar phase}. 
We explain the details of the model below. 

%%%%%%%%%%%%%%%%%
%%%%%%%%%%%%%%%%%
\subsection{Collapsing phase}
%%%%%%%%%%%%%%%%%
%%%%%%%%%%%%%%%%%

The gravitational collapse of a spherically symmetric 
star composed of homogeneous dust is described by the Oppenheimer-Snyder solution. 
The metric inside the star is given by Eq.~\eqref{RW-metric} with $\varSigma(\chi)=\sin\chi$. 
The surface of the dust sphere is $\chis=\chi_{\rm i}=$ a positive constant, and we have
\begin{align}
a&=\frac{GM}{\sin^3\chii}\left(1+\cos\eta\right), \\
\tau&=\frac{GM}{\sin^3\chii}(\eta+\sin\eta),
\end{align}
where $M$ is a positive constant representing the gravitational mass of the star, 
and the domain of $\eta$ is restricted to $\eta\geq0$.  
In this model, a black hole forms when $\rs=\as \sin\chi_{\rm i}=2GM$ is satisfied, or equivalently,  
\begin{equation}
\eta=\eta_{\rm bh}:=\pi-2\chi_{\rm i} \label{eta-bh}
\end{equation}
is satisfied. Note that $\chii$ should be less than $\pi/2$ so that $\eta_{\rm bh}$ is positive. 

%%%%%%%%%%%%%%%%%
%%%%%%%%%%%%%%%%%
\subsection{Parameters to determine the duration and characteristic radius of each phase}
%%%%%%%%%%%%%%%%%
%%%%%%%%%%%%%%%%%

As mentioned, we assume that the gravitational collapse stops before the formation of a black hole at $\eta=\eta_{\rm bh}$, 
and the star eventually becomes a gravastar.  
We represent the conformal time $\etaf$ at which the star becomes the static gravastar by introducing a small positive parameter $\epsilon$ as 
\begin{equation}
\etaf=\eta_{\rm bh}-\epsilon^2. \label{eta-f}
\end{equation}
By introducing a positive parameter $\beta$, the conformal time $\etab$ at which the collapsing motion begins slowing down denotes 
\begin{equation}
\etab=\etaf-\epsilon^{2\beta}=\eta_{\rm bh}-\epsilon^2-\epsilon^{2\beta}
\end{equation}
so that $\etab<\etaf$. Thus, the collapsing phase, the slowing-down phase and the gravastar phase are $0\leq\eta<\etab$, 
$\etab\leq\eta<\etaf$, and $\etaf\leq\eta$, respectively.  
 
The areal radius of the star at the beginning of the slowing-down phase, $\rs=\rb$, is given as
\begin{equation}
\rb=\frac{GM}{\sin^2\chii}(1+\cos\etab) 
\end{equation}
Note that the areal radius of the gravastar, $\rf$, is also a free parameter of the model. 
By introducing a parameter $\alpha$ restricted to $0<\alpha<1$, we represent it as
\begin{align}
\rf&=\frac{GM}{\sin^2\chii}\left\{1+\cos\left[\etab+(1-\alpha)\epsilon^{2\beta}\right]\right\} \nonumber \\
&=\frac{GM}{\sin^2\chii}\left[1+\cos(\etaf-\alpha\epsilon^{2\beta})\right] \label{alpha-def}
\end{align}
so that $\rb>\rf$. Note that both $\rb$ and $\rf$ are larger than the gravitational radius $2GM$ and approach $2GM$ 
in the limit of $\epsilon\rightarrow0$.

%%%%%%%%%%%%%%%%%
%%%%%%%%%%%%%%%%%
\subsection{Gravastar phase}
%%%%%%%%%%%%%%%%%
%%%%%%%%%%%%%%%%%

Before mentioning the detail of the slowing-down phase, we describe the gravastar as a final product in the present model. 
The gravastar is a horizonless ultra-compact object composed of the dark energy with the equation of state $p=-\rho<0$, or equivalently, 
the positive cosmological constant, which is  
enclosed by an infinitesimally thin crust\cite{MM2004,VW2004}. In the present model, 
the equation of state of the crust is assumed to be determined through Darmois-Israel junction condition (see Appendix B)\cite{Israel:1966}.
Due to the assumption of the spherical symmetry, the inside of the gravastar is described by the de Sitter geometry. 
The scale factor of the de Sitter spacetime in the  coordinate system of Eq.~\eqref{RW-metric} with $\varSigma=\sin\chi$ 
is given as
\begin{equation}
a(\eta)=\frac{1}{H\sin\left(\eta-\etac\right)}, \label{a-eta-relation}
\end{equation}
where $\etac<\eta<\pi+\etac$, $\etac$ is a constant determined later, and $H$ is the so called Hubble parameter of the de Sitter spacetime 
which is related to the positive cosmological constant $\Lambda$ by
\begin{equation}
H=\sqrt{\frac{\Lambda}{3}}. 
\end{equation}
The final state of the gravastar is static, and its areal radius is equal to $\rf$ expressed as Eq.~\eqref{alpha-def}. 
Due to the relation $\rs=\as\sin\chis$, the relation 
$\rf=\as\sin\chis$ should be satisfied for the static gravastar. Hence, we have 
\begin{equation}
\chis= \arcsin\left[H\rf\sin(\etas-\etac)\right] \label{chi-asympt}
\end{equation}
in the final state. Note that $\chis$ is not constant. It should also be noted that an inequality
\begin{equation}
\rf<H^{-1}
\end{equation}
should hold so that the timelike Killing vector in the de Sitter spacetime is tangent to the surface of the gravastar.  

In the original scenario of the gravastar, $H^{-1}$ is almost equal to $2GM$\cite{MM2004}. Thus, for simplicity, we assume here 
\begin{equation}
H^2\rf^2=\frac{2GM}{\rf} .
\label{Hubble}
\end{equation}
By using formulae given in Appendix B, we find 
that this condition leads to the equation of state of the crust whose surface energy density 
$S_{(0)(0)}$ vanishes and tangential stresses $S_{(2)(2)}$ and $S_{(3)(3)}$ is given as 
\begin{equation}
S_{(2)(2)}=S_{(3)(3)}=\frac{3M}{8\pi\rf^2\sqrt{f(\rf)}}.
\end{equation} 
The crust of the gravastar does not satisfy reasonable energy conditions, e.g., the dominant energy condition(see Ref.~\cite{Wald}). 
However, the possibility of such a gravastar model may not be excluded, since the gravastar model is a proposal by assuming 
some unknown framework beyond Einstein's theory, in which the 
effective stress-energy-momentum tensor does not necessarily satisfy the reasonable energy conditions.

Here, we determine $\etac$.  From Eqs. \eqref{eta-bh} and \eqref{eta-f}, the formation time of the static gravastar is equal to $\etaf=\pi-2\chii-\epsilon^2$. 
Equation \eqref{a-eta-relation} implies that  $\as$ diverges in the limit of $\eta-\etac\rightarrow\pi$, and hence 
the static gravastar appears in $\etaf < \eta <\pi +\etac$, or equivalently, $\pi-(2\chii+\epsilon^2)<\eta <\pi+\etac$. 
As a result, we find that the inequality
\begin{equation}
\etac>-(2\chii+\epsilon^2) \label{etac-constraint}
\end{equation}
should hold. 
The variation of the areal radius of the star in the slowing-down phase, $\rf-\rb$, is approximately given as
\begin{equation}
\rf-\rb\simeq-\frac{GM\sin(2\chii+\epsilon^2)}{\sin^2\chis}(1-\alpha)\epsilon^{2\beta}.
\end{equation}
We impose that the variation of $\chis$ in the slowing-down phase is the order of max$\left[\epsilon^2,\epsilon^{2\beta}\right]$:
\begin{equation}
\sin\chis|_{\etas=\etaf}-\sin\chii=H\rf\sin(\etaf-\etac)-\sin\chii= {\cal O}\left(\rm{max}\left[\epsilon^2,\epsilon^{2\beta}\right]\right)<0,
\label{etac-reason}
\end{equation}
where, since $\chis$ is a decreasing function of $\etas$ in the gravastar phase, we have also imposed the last inequality, i.e., 
the variation of $\chis$ is negative.  Thus, for example, we put $\etac$ as
\begin{equation}
\etac=-\left(\chii+\frac{3}{2}\epsilon^2\right), \label{eta-c}
\end{equation}
so that Eq.~\eqref{etac-reason} holds (see Eq.~\eqref{chi-G} in the case of $0<\beta<1$), where we assume $\chii>\epsilon^2$.

%%%%%%%%%%%%%%%%%
%%%%%%%%%%%%%%%%%
\subsection{Slowing-down phase before gravastar formation}
%%%%%%%%%%%%%%%%%
%%%%%%%%%%%%%%%%%

We determine the behavior of the star for the slowing-down phase, $\etab<\eta<\etaf$. 
We assume both the radial coordinates $\chis$ and $\rs$ of the surface of the star are $C^3$ functions of $\etas$. 
Then, we represent $\chis$ as 
\begin{equation}
\sin\chis=
\begin{cases}
\sin\chii &\text{for $0<\etas\leq\etab$} \\
\sin\chii+\displaystyle{\sum_{n=0}^3}c_n(\etas-\etab)^{n+4}      &\text{for $\etab< \etas < \etaf$} \\
H\rf\sin(\etas-\eta_{\rm c}) &\text{for $\etaf\leq\etas<\etac+\pi$}
\end{cases}
~.\label{chis}
\end{equation}
where putting 
\begin{align}
S_\chi&:=H\rf\sin(\etaf-\eta_{\rm c}), \\
C_\chi&:=H\rf\cos(\etaf-\eta_{\rm c}),
\end{align}
the coefficients $c_n$ are given as
\begin{align}
c_0&=\frac{1}{6\epsilon^{8\beta}}\left[210(S_\chi-\sin\chii)-90C_\chi\epsilon^{2\beta} -15S_\chi\epsilon^{4\beta} +C_\chi\epsilon^{6\beta}\right],\\
c_1&=\frac{1}{2\epsilon^{10\beta}}\left[-168(S_\chi-\sin\chii)+78C_\chi\epsilon^{2\beta} +14S_\chi\epsilon^{4\beta} -C_\chi\epsilon^{6\beta}\right],\\
c_2&=\frac{1}{2\epsilon^{12\beta}}\left[140(S_\chi-\sin\chii)-68C_\chi\epsilon^{2\beta} -13S_\chi\epsilon^{4\beta} +C_\chi\epsilon^{6\beta}\right],\\
c_3&=\frac{1}{6\epsilon^{14\beta}}\left[-120(S_\chi-\sin\chii)+60C_\chi\epsilon^{2\beta} +12S_\chi\epsilon^{4\beta} -C_\chi\epsilon^{6\beta}\right].
\end{align}

Since the world line of the surface of the star should be timelike, 
 \begin{equation}
\left|\frac{d\chis}{d\etas}\right|<1 \label{v-constraint}
\end{equation}
should hold. As shown below, this condition constrains the parameter $\beta$. 
Since the period $\etab<\etas<\etaf=\etab+\epsilon^{2\beta}$ is the slowing-down phase, 
we introduce a normalized time coordinate $\gamma$ for this phase, which is defined as
\begin{equation}
\etas=\etab+\gamma\epsilon^{2\beta},
\end{equation}
where $\gamma$ increases from $0$ to $1$.  Then, by differentiating Eq.~\eqref{chis}, we have 
\begin{align}
\cos\chis\frac{d\chis}{d\etas}&=\gamma^3\biggl\{140(1-\gamma)^3(S_\chi-\sin\chii)\epsilon^{-2\beta}
+\left(70\gamma^3-204\gamma^2+195\gamma-60\right)C_\chi \nonumber \\
&-(1-\gamma)\left(14\gamma^2-25\gamma+10\right)S_\chi\epsilon^{2\beta}
+\frac{1}{6}(1-\gamma)\left(7\gamma^2-11\gamma+4\right)C_\chi\epsilon^{4\beta}\biggr\} \label{cos-dchi-deta}
\end{align}
We are only interested in the case of $0<\epsilon\ll1$. In the limit of  $\epsilon\rightarrow0+$, we have
\begin{align}
\cos\chis&\longrightarrow \cos\chii, \\
C_\chi&\longrightarrow-\cos\chii, \\
S_\chi&\longrightarrow\sin\chii,
\end{align}
and
\begin{equation}
(S_\chi-\sin\chi)\epsilon^{-2\beta}\longrightarrow 
\begin{cases}
-\dfrac{1}{2}\alpha\cos\chii                      &\text{for $0<\beta<1$} \\
-\left(1+\dfrac{1}{2}\alpha\right)\cos\chii     &\text{for $\beta=1$} \\
-\cos\chii~ \displaystyle{\lim_{\epsilon\rightarrow0}}\epsilon^{-2(\beta-1)}=-\infty    &\text{for $\beta>1$}
\end{cases}
\label{1st-term}
\end{equation}
Hence, $0<\beta\leq1$ should be imposed so that Eq.~\eqref{v-constraint} holds for very small positive $\epsilon$. 
Further careful investigation shows that $\beta=1$ should be excluded (see Appendix C).  Hence, hereafter, we assume 
\begin{equation}
0<\beta<1.
\end{equation}

We represent $\rs$ as
\begin{equation}
\rs=
\begin{cases}
\dfrac{GM}{\sin^2\chii}(1+\cos\eta) &\text{for $0<\etas\leq\etab$} \\
\rf+\displaystyle{\sum_{n=0}^3} C_n(\etas-\etaf)^{n+4}      &\text{for $\etab< \etas < \etaf$} \\
\rf &\text{for $\etaf\leq\etas<\etac+\pi$}
\end{cases}
~.\label{rs}
\end{equation}
where putting
\begin{align}
S_r&:=\frac{GM}{\sin^2\chii}\sin\etab, \\
C_r&:=\frac{GM}{\sin^2\chii}\cos\etab,
\end{align}
the coefficients $C_n$ are given as
\begin{align}
C_0&=\frac{1}{6\epsilon^{8\beta}}\left[210(\rb-\rf)-90S_r\epsilon^{2\beta} -15C_r\epsilon^{4\beta} +S_r\epsilon^{6\beta}\right],\label{C0}\\
C_1&=\frac{1}{2\epsilon^{10\beta}}\left[168(\rb-\rf)-78S_r\epsilon^{2\beta} -14C_r\epsilon^{4\beta} +S_r\epsilon^{6\beta}\right],\\
C_2&=\frac{1}{2\epsilon^{12\beta}}\left[140(\rb-\rf)-68S_r\epsilon^{2\beta} -13C_r\epsilon^{4\beta} +S_r\epsilon^{6\beta}\right],\\
C_3&=\frac{1}{6\epsilon^{14\beta}}\left[120(\rb-\rf)-60S_r\epsilon^{2\beta} -12C_r\epsilon^{4\beta} +S_r\epsilon^{6\beta}\right].
\end{align}

The scale factor $\as$ is given by
\begin{equation}
\as=\frac{\rs}{\sin\chis}. \label{sfactor}
\end{equation}

Since the radial coordinates $\chis$ and $\rs$ and the scale factor $a$ are explicitly expressed 
as functions of $\etas$ as in Eqs.~\eqref{chis}, \eqref{rs} and \eqref{sfactor}, 
we represent the derivatives of them with respect to $\tau$ by using their derivatives with respect to $\etas$. 
From the normalization condition $a^2\left(\dot{\eta}_{\rm s}^2-\dchis^2\right)=1$, we have
\begin{equation}
\dot{\eta}_{\rm s}=\frac{1}{\as\sqrt{1-\left(\dfrac{d\chis}{d\etas}\right)^2}}. \label{etas-dot}
\end{equation}
By using this relation, we obtain
\begin{align}
\dchis&=\dot{\eta}_{\rm s}\frac{d\chis}{d\etas}=\frac{1}{\as\sqrt{1-\left(\dfrac{d\chis}{d\etas}\right)^2}}\dfrac{d\chis}{d\etas}, \label{chi-dot}\\
\ddot{\chi}_{\rm s}&=\frac{1}{\as^2\left[1-\left(\dfrac{d\chis}{d\etas}\right)^2\right]^2}\dfrac{d^2\chis}{d\etas^2}
-\frac{1}{\as^3\left[1-\left(\dfrac{d\chis}{d\etas}\right)^2\right]}\frac{d\as}{d\etas}\frac{d\chis}{d\etas}, \label{chi-ddot}\\
\drs&=\dot{\eta}_{\rm s}\frac{d\rs}{d\etas}=\frac{1}{\as\sqrt{1-\left(\dfrac{d\chis}{d\etas}\right)^2}}\dfrac{d\rs}{d\etas}, \\
\ddot{r}_{\rm s}&=\frac{1}{\as^2\left[1-\left(\dfrac{d\chis}{d\etas}\right)^2\right]^2}
\left[\dfrac{d^2\rs}{d\etas^2}
+\dfrac{d\chis}{d\etas}\left(\dfrac{d\rs}{d\etas}\dfrac{d^2\chis}{d\etas^2}-\dfrac{d^2\rs}{d\etas^2}\dfrac{d\chis}{d\etas}\right)\right] \nonumber \\
&-\frac{1}{\as^3\left[1-\left(\dfrac{d\chis}{d\etas}\right)^2\right]}\frac{d\as}{d\etas}\frac{d\rs}{d\etas}. \label{rs-ddot}
\end{align}
From Eq.~\eqref{sfactor}, we have
\begin{align}
\frac{d\as}{d\etas}&=\as\left(\frac{1}{\rs}\frac{d\rs}{d\etas}-\cot\chis \frac{d\chis}{d\etas}\right), \label{da-deta} 
\end{align}

As shown in Appendix E, the dominant energy condition does not hold in this phase for $0<\epsilon\ll1$.

%%%%%%%%%%%%%%%%%
%%%%%%%%%%%%%%%%%
\subsection{Time coordinate outside the star}
%%%%%%%%%%%%%%%%%
%%%%%%%%%%%%%%%%%

In order to see the $\eta$-dependence of $\ts$, we need to solve the following differential equation: 
\begin{equation}
\frac{d\ts}{d\etas}=\frac{d\tau}{d\etas}\dot{t}_{\rm s}
=\frac{1}{f(\rs)}\sqrt{\left(\frac{d\rs}{d\etas}\right)^2+\as^2 f(\rs)\left[1-\left(\frac{d\chis}{d\etas}\right)^2\right]}.
\label{dtdeta}
\end{equation}
In the dust collapsing phase, $0<\etas<\etab$, this equation becomes
\begin{equation}
\frac{d\ts}{d\etas}=\frac{GM\cos\chii}{\sin^3\chii}\left(1+2\sin^2\chii+\cos\etas+\frac{4\sin^4\chii}{\cos\etas+\cos2\chii}\right).
\end{equation}
This differential equation can be analytically solved, and we have
\begin{equation}
\ts=\frac{GM\cos\chii}{\sin^3\chii}\left[
(1+2\sin^2\chii)\etas+\sin\etas+\frac{2\sin^3\chii}{\cos\chii}
\ln\left|\frac{\cot\chii+\tan\dfrac{\etas}{2}}{\cot\chii-\tan\dfrac{\etas}{2}}\right|
\right], \label{collapse-time}
\end{equation}
where we have chosen the integration constant so that $\ts$ vanishes at $\etas=0$. 

In the static gravastar phase, $\etas>\etaf$, Eq.~\eqref{dtdeta} becomes
\begin{equation}
\frac{d\ts}{d\etas}=\frac{1}{H\sin\left(\etas-\etac\right)\sqrt{1-H^2\rf^2\sin^2\left(\etas-\etac\right)}}.
\end{equation}
This differential equation can also be analytically integrated and we obtain
\begin{align}
\ts&=\frac{1}{2H}
\ln\left|\frac{H\rf\sin^2(\etas-\etac)+1-\cos(\etas-\etac)\sqrt{1-H^2\rf^2\sin^2\left(\etas-\etac\right)}}
{H\rf\sin^2(\etas-\etac)-1-\cos(\etas-\etac)\sqrt{1-H^2\rf^2\sin^2\left(\etas-\etac\right)}}\right| +\tc, \label{gravastar-t}
\end{align}
where $\tc$ is an integration constant. For the slowing-down phase, $\etab<\etas<\etaf$, we need to numerically integrate Eq.~\eqref{dtdeta}. 
In order that the time coordinate $\ts$ of the surface of the star is continuous at $\etas=\etaf$, the integration constant $\tc$ should satisfy 
\begin{align}
\tc&=\int_{\etab}^{\etaf}
\frac{1}{f(\rs)}\sqrt{\left(\frac{d\rs}{d\etas}\right)^2+\as^2 f(\rs)\left[1-\left(\frac{d\chis}{d\etas}\right)^2\right]}
d\etas
\nonumber \\
&+\frac{GM\cos\chii}{\sin^3\chii}\left[
(1+2\sin^2\chii)\etab+\sin\etab+\frac{2\sin^3\chii}{\cos\chii}
\ln\left|\frac{\cot\chii+\tan\dfrac{\etab}{2}}{\cot\chii-\tan\dfrac{\etab}{2}}\right|\right] \nonumber \\
&-\frac{1}{2H}
\ln\left|\frac{H\rf\sin^2(\etaf-\etac)+1-\cos(\etaf-\etac)\sqrt{1-H^2\rf^2\sin^2(\etaf-\etac)}}
{H\rf\sin^2(\etaf-\etac)-1-\cos(\etaf-\etac)\sqrt{1-H^2\rf^2\sin^2(\etaf-\etac)}}\right| .
\end{align}

%%%%%%%%%%%%%%%%%%%%%%%%%%%%%
%%%%%%%%%%%%%%%%%%%%%%%%%%%%%
%%%%%%%%%%%%%%%%%%%%%%%%%%%%%
\section{An example and classification of radial null geodesics}
%%%%%%%%%%%%%%%%%%%%%%%%%%%%%
%%%%%%%%%%%%%%%%%%%%%%%%%%%%%
%%%%%%%%%%%%%%%%%%%%%%%%%%%%%

%%%%%%%%%%%%%
%%%%%%%%%%%%%
\begin{figure}
\begin{center}
\includegraphics[width=0.85\textwidth]{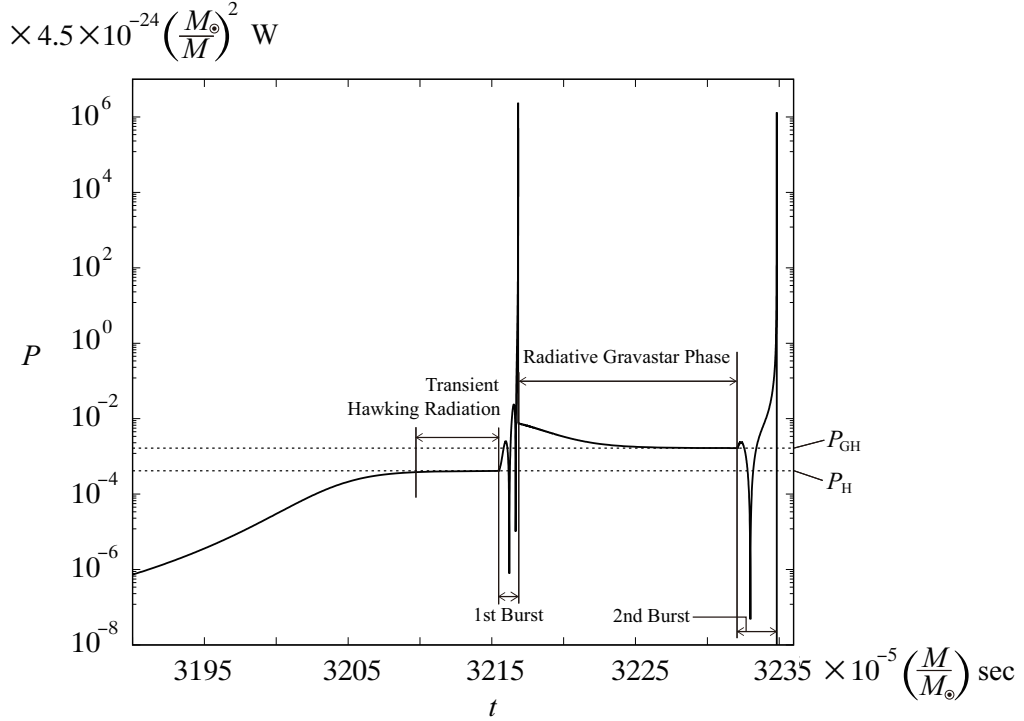}
\caption{\label{r-power}
The radiation power $P$ in the case of $\chii=0.1$, $\alpha=\beta=0.5$, and $\epsilon=10^{-4}$ is depicted as a function of time $t$. 
$P_{\rm H}$ and $P_{\rm GH}$ are the values of the radiation power of the Hawking radiation with the gravitational mass $M$ and 
that of the Gibbons-Hawking radiation with the Hubble constant $H$, respectively. 
}
\end{center}
\end{figure}
%%%%%%%%%%%%%
%%%%%%%%%%%%%

In Fig.~\ref{r-power}, we show an example of the the radiation power detected in the asymptotic region $r\gg 2GM$ as a function of time $t$; 
$\chii=0.1$, $\alpha=\beta=0.5$, and $\epsilon=10^{-4}$ are assumed. We numerically checked that the world line of the surface 
of the collapsing object is timelike in this case. 
The initial radius at $\etas=0$ is equal to $2GM/\sin^2\chii\simeq 200GM$. From this figure, we find that there are three characteristic 
periods divided by two bursts of the quantum particle creation. 
In the first period, the radiation power $P$ increases from almost zero and 
eventually becomes the value of the so-called Hawking radiation 
\begin{equation}
P_{\rm H}=\frac{1}{48\pi (4GM)^2}.
\end{equation}
This value is kept until the 1st burst.  Since the adiabatic condition $|\kappa'|\ll\kappa^2$ holds at this stage,  
the spectrum is thermal one with the so-called Hawking temperature 
\begin{equation}
k_{\rm B}T_{\rm H}=\frac{1}{8\pi GM}.
\end{equation} 
Thus, following Refs.~\cite{Harada-CM,Kokubu-H}, we call this radiation the transient Hawking radiation. 
In the period between the first and second bursts, 
the radiation power $P$ becomes small but does not vanish; 
later, we will discuss this non-vanishing radiation power in detail. 
After the second burst, the  radiation power $P$ vanishes completely.

%%%%%%%%%%%%%
%%%%%%%%%%%%%
\begin{figure}
\begin{center}
\includegraphics[width=0.75\textwidth]{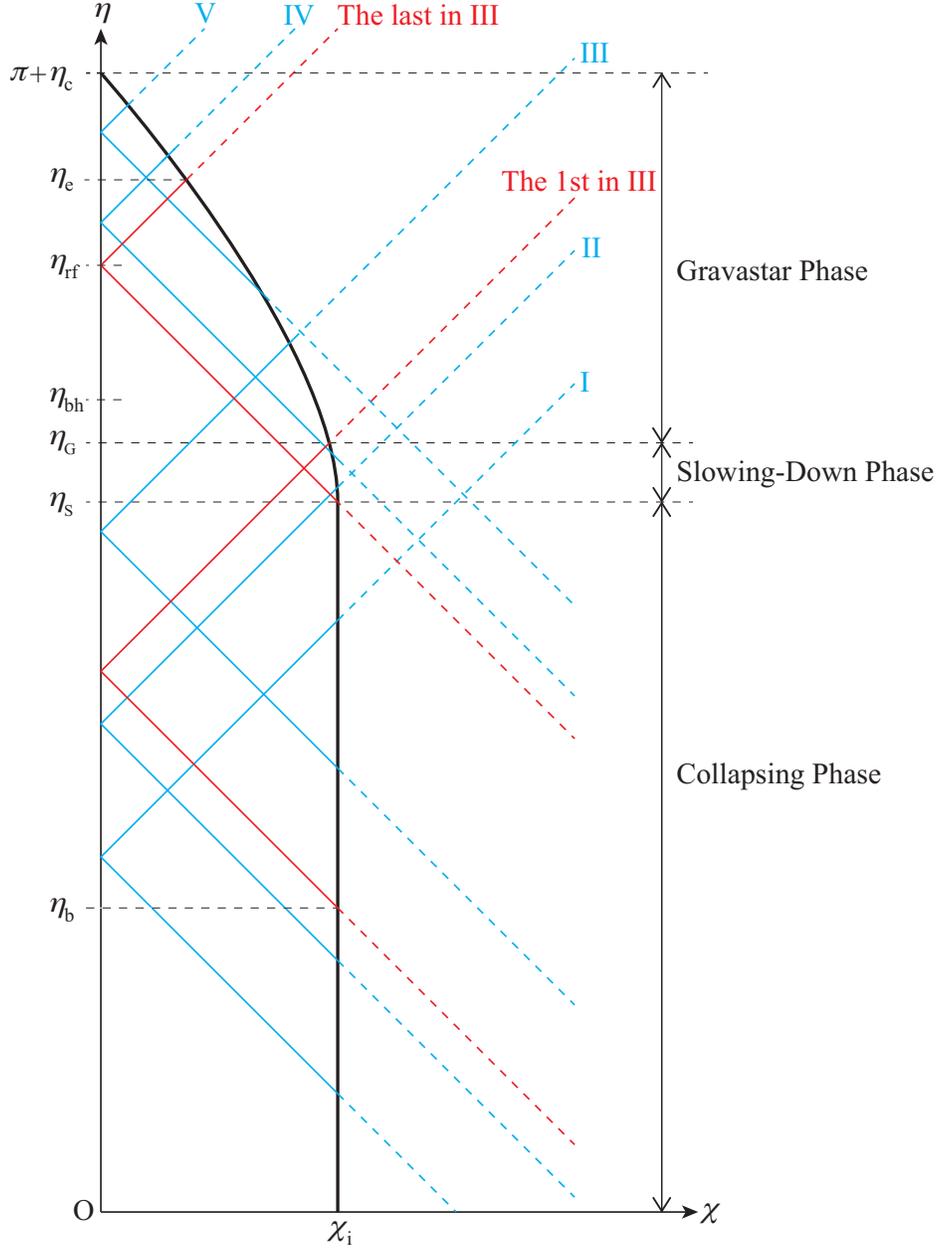}
\caption{\label{spacetime}
The schematic spacetime diagram by using the coordinates $(\eta,\chi)$ inside the star is depicted. The thick solid black curve is the 
world line of the surface of the star. Hence, only the left-hand-side domain of this curve is covered by this coordinates. 
Blue lines represent the radial null geodesics, each of which is a typical one of Class I, II, III, IV and V.  
The two red lines are the first and last radial null geodesics in Class III. Note that the areal radius $\rs$ of the surface of the star is equal to the constant $\rf$ 
in the gravastar phase. 
}
\end{center}
\end{figure}
%%%%%%%%%%%%%
%%%%%%%%%%%%%

As mentioned, we adopt the $S$-wave approximation in order to estimate the radiation power of quantum particle creation. 
In this approximation scheme, the flux of particles detected at the asymptotic region comes from the nontrivial deformation of the phase of  
the spherically symmetric mode function propagating from the past null infinity to the future null infinity. 
Figure \ref{spacetime} depicts the spacetime diagram by using the coordinates $(\eta,\chi)$. 
The black solid curve is the world line of the surface of the star. Blue lines represent the radial null geodesics along which the 
spherically symmetric mode function propagate: each null geodesic is categorized into the following five classes:
\begin{itemize} 
\item Class I: null geodesics categorized into this class enter and come out from the star in the collapsing phase. 
\item Class II: null geodesics categorized into this class enter the star in the collapsing phase and come out from the star 
in the slowing-down phase. 
\item Class III: null geodesics categorized into this class enter the star in the collapsing phase and come out from the star 
in the gravastar phase. 
\item Class IV: null geodesics categorized into this class enter the star in the slowing-down phase and come out from the star 
in the gravastar phase. 
\item  Class V: null geodesics categorized into this class enter and come out from the star in the gravastar phase. 
\end{itemize} 
The mode function along the null geodesic in Class I causes the transient Hawking radiation if the null comes out from the star 
in the very late stage of the collapsing phase.
The mode function propagating along the null geodesics in Class II causes the 1st burst, i.e, the post-Hawking burst named in Ref.~\cite{Harada-CM}. 
The mode function along the null geodesics in Class III causes the non-vanishing radiation. 
The mode function along the null geodesics in Class IV causes the 2nd burst. 
The mode function along the null geodesics in Class V causes no particle creation, since these null geodesics only go through 
the static domain. 

The occurrence of the two bursts of the quantum particle creation is a common feature to both the present 
study and precedent ones\cite{Harada-CM,Okabayashi-HN}. 
By contrast, the non-vanishing radiation in the period between two bursts is the characteristic of the present model, since there is no radiation 
in this period in the models investigated in the precedent studies.  The gravastar has already formed in the causal past of the asymptotic 
observer after the 1st burst, and therefore this radiation can be attributed to the gravastar formation. We call this period 
{\it the radiative gravastar phase}. Details of this phase will be discussed in the next subsection.  

The complicated temporal variation of the radiation power $P$ during the two bursts is also characteristic in the present model. 
This temporal variation merely comes from a bit complicated behavior of the radial coordinate of the surface of the star, 
$\chis$ and $\rs$, especially non-monotonicity of the dependence of $\chis$ on the conformal time $\etas$. 
By differentiating Eqs.~\eqref{chis} and \eqref{rs} with respect to $\etas$ once or twice and taking a limit of $\epsilon\rightarrow0$, we have
\begin{align}
\frac{d\chis}{d\etas}\longrightarrow&-\gamma^3\left[1+3(1-\gamma)+6(1-\gamma)^2-70(1-\alpha)(1-\gamma)^3\right], \label{dchi-deta}\\
\frac{d\rs}{d\etas}\longrightarrow&-2GM(1-\gamma)^3\left[70(1-2\alpha)\gamma^3+6\gamma^2+3\gamma+1\right]\cot\chii ,\\
\epsilon^{2\beta}\frac{d^2\chis}{d\etas^2}\longrightarrow& 5\gamma^2\left[\alpha-6(2-7\alpha)(1-\gamma)^2+84(1-\alpha)(1-\gamma)^3\right], \\
\epsilon^{2\beta}\frac{d^2\rs}{d\etas^2}\longrightarrow&60GM\gamma^2(1-\gamma)^2\left[7(1-2\alpha)\gamma+7\alpha-3\right]\cot\chii.\label{ddrs-deta-lim}
\end{align}
for the slowing-down phase $\etab<\etas<\etaf$. 
Substituting these results into Eqs.~\eqref{chi-dot}--\eqref{da-deta} and 
using Eqs.~\eqref{kappa-out}, \eqref{A-rep}--\eqref{ABCD-def}, we find that the amplitude of the radiation power $P$ of the burst 
is proportional to $\epsilon^{-4\beta}$ for $1-\gamma\gg\epsilon^{2\beta}$, 
in the case that the mode function propagates along the null in Class II or Class IV.

%%%%%%%%%%%%%%%%%%%%%%%%%%%%%
%%%%%%%%%%%%%%%%%%%%%%%%%%%%%
%%%%%%%%%%%%%%%%%%%%%%%%%%%%%
\section{The radiative gravastar phase}
%%%%%%%%%%%%%%%%%%%%%%%%%%%%%
%%%%%%%%%%%%%%%%%%%%%%%%%%%%%
%%%%%%%%%%%%%%%%%%%%%%%%%%%%%
As mentioned, in the radiative gravastar phase, 
the mode function propagates along radial null geodesics categorized into Class III. 
We investigate the quantity $\kappa$ associated to the null geodesics in Class III in this subsection. 

%%%%%%%%%%%%%%%%%
%%%%%%%%%%%%%%%%%
\subsection{Calculation of $\kappa$}
%%%%%%%%%%%%%%%%%
%%%%%%%%%%%%%%%%%
The first radial null geodesic of Class III  comes out from the star at $\etas=\etaf$, and hence is described by  
$\eta-\etaf=\chi-\chis|_{\etas=\etaf}$. It intersects the origin $\chi=0$ at $\eta=\etaf-\chis|_{\etas=\etaf}$ and 
is the ingoing null $\eta-\etaf+\chis|_{\etas=\etaf}=-\chi$ before this moment. The conformal time 
at which the first radial null geodesic enters the star is denoted by $\etabgn$. 
Since the star is in the collapsing phase with $\chis=\chii$ at this moment, we have 
\begin{equation}
\etabgn=\etaf-\chis|_{\etas=\etaf}-\chii\simeq \pi-4\chii. \label{eta-begin}
\end{equation}

By contrast, the last radial null geodesic in Class III enters the star at $\etas=\etab$. 
The ingoing radial null arriving at the surface of the star at $\eta=\etab$ is represented as 
$\eta-\etab=-\chi+\chii$ since $\chis|_{\etas=\etab}=\chii$ holds. It will arrive at the center $\chi=0$ at
\begin{equation}
\eta=\etar:=\etab+\chis|_{\etas=\etab}\simeq\pi-\chii-\epsilon^{2\beta}, \label{etar}
\end{equation}
and then becomes the outgoing radial null represented as $\eta-\etar=\chi$. The radial null which has become outgoing again crosses 
the surface of the star, and we denote the conformal time at this moment by $\etaend$. 
Since $\etaend-\etar=\chis|_{\etas=\etaend}$ holds, Eq.~\eqref{chi-asympt} leads to, 
\begin{equation}
\sin(\etaend-\etar)=H\rf\sin(\etaend-\etac). \label{eta-out-eq}
\end{equation}
Rewriting the left hand side of this equation as
\begin{equation}
\sin\left[\etaend-\etac+(\etac-\etar)\right]=\sin(\etaend-\etac)\cos(\etac-\etar)+\sin(\etac-\etar)\cos(\etaend-\etac),
\end{equation}
we obtain 
\begin{equation}
\left[\cos(\etaend-\etac)-H\rf\right]\sin(\etaend-\etac)=-\sin(\etac-\etar)\cos(\etaend-\etac).
\end{equation}
By taking the square of both sides of this equation, we have
\begin{equation}
\sin^2(\etaend-\etac)=\frac{\sin^2(\etac-\etar)}{1-2H\rf\cos(\etac-\etar)+H^2\rf^2}.
\end{equation}
Since $\sin(\etaend-\etac)>0$ holds and, as shown below, $\sin(\etac-\etar)<0$ is satisfied, we have
\begin{equation}
\sin(\etaend-\etac)=-\frac{\sin(\etac-\etar)}{\sqrt{1-2H\rf\cos(\etac-\etar)+H^2\rf^2}}. \label{sin-out}
\end{equation}
By using Eqs.~\eqref{eta-c} and \eqref{etar}, we obtain
\begin{equation}
\etac-\etar\simeq-\left(\pi-\epsilon^{2\beta}\right),
\end{equation}
and hence we have
\begin{equation}
\sin(\etac-\etar)\simeq-\epsilon^{2\beta}~~~{\rm and}~~~\cos(\etac-\etar)\simeq -1+\frac{1}{2}\epsilon^{4\beta}.
\end{equation}
By these equations and Eq.~\eqref{sin-out}, we have
\begin{equation}
\sin(\etaend-\etac)\simeq \frac{1}{2}\epsilon^{2\beta}.
\end{equation}
Since $\etaend-\etac\simeq \pi$ holds, 
\begin{equation}
\etaend-\etac\simeq \pi-\frac{1}{2}\epsilon^{2\beta}  \label{etaend-etac}
\end{equation}
is obtained. 

It is easy to obtain the following results for the null which enters the star in the collapsing phase. Since 
$\dot{\chi}_{\rm s}=0=\ddot{\chi}_{\rm s}$ holds in this phase, we have
\begin{align}
\frac{\Din}{\Bin}&=-\frac{GM}{\rs\sin\chii(\cos\chii-\drs)}+\frac{\drs}{\sin\chii} \nonumber \\
&=-\frac{1}{2}\left[\frac{1-f(\rs)}{f(\rs)}\left(1-\sqrt{1-\frac{f(\rs)}{\cos^2\chii}}\right)+2\sqrt{1-\frac{f(\rs)}{\cos^2\chii}}\right]\cot\chii
\label{DB-ratio-0}
\end{align}
where we have used the fact that 
\begin{equation}
\dot{r}_{\rm s}^2=\cos^2\chii-f(\rs)
\end{equation}
holds in the collapsing phase. 
The first radial null geodesic enters the star with $\rs=\rs|_{\etas=\etabgn}\simeq 8GM\cos^2\chii$, 
which is assumed to be in the collapsing phase, i.e., $0<\etabgn<\etab$. 
Because of $0<\etabgn \simeq \pi-4\chii+\alpha\epsilon^{2\beta}/2$ (see Eqs.~\eqref{eta-begin} and \eqref{chi-G}), we impose 
a stringent restriction on $\chii$ more than $0<\chii<\pi/2$ mentioned below Eq.~\eqref{eta-bh} as
\begin{equation}
0<\chii \leq \pi/4. 
\end{equation}
Hence we have 
\begin{equation}
\left.\frac{\Din}{\Bin}\right|_{\rm 1st} \simeq -\frac{4\cos^2 2\chii+2\cos2\chii+1}{2\sin2\chii\left(2\cos2\chii+1\right)}.\label{DB-ratio-1st}
\end{equation}
By contrast, since the last radial null geodesic enters the star with $\rs=\rb\simeq 2GM$ (see Eq.~\eqref{rb-app}), we have
\begin{equation}
\left.\frac{\Din}{\Bin}\right|_{\rm last} \simeq -\frac{2\cos2\chii+3}{2\sin2\chii}.\label{DB-ratio-last}
\end{equation}
%\Red{where we have used the fact $\dot{r}_{\rm s}\simeq -\cos\chii<0$. Note that , as mentioned below Eq.~\eqref{eta-bh}, $0<\chii<\pi/2$ holds. }

In the gravastar phase, $\dot{r}_{\rm s}=0=\ddot{r}_{\rm s}$ holds, whereas we have 
\begin{align}
\dot{\chi}_{\rm s}&=-\frac{\rf}{\as\sqrt{1-H^2\rf^2}}\sqrt{H^2-\frac{1}{\as^2}}, \\
\ddot{\chi}_{\rm s}&=\frac{\rf}{\as(1-H\rf)}\left(H^2-\frac{2}{\as^2}\right)\sqrt{1-\frac{\rf^2}{\as^2}}.
\end{align}
By using these results, we obtain
\begin{align}
\Aout&=\frac{1}{\as}\left(\sqrt{1-\frac{\rf^2}{\as^2}}+H\rf\sqrt{1-\frac{1}{H^2\as^2}}\right)=\frac{\sin\chis}{\rf}\left(\cos\chis+\sqrt{\cos^2\chis-f(\rf)}\right), \\
\Cout&=H\sqrt{1-\frac{1}{H^2\as^2}}\left(\sqrt{1-\frac{\rf^2}{\as^2}}-H\rf\sqrt{1-\frac{1}{H^2\as^2}}\right)
+H^2\rf\left(1-\frac{2}{H^2\as^2}\right) \nonumber \\
&=\frac{1}{\rf}\sqrt{\cos^2\chis-f(\rf)}\left(\cos\chis-\sqrt{\cos^2\chis-f(\rf)}\right)+\frac{\cos2\chis-f(\rf)}{\rf}.
\end{align}
where we have used $\as =\rf/\sin\chis$. Since $0<f(\rf)=1-H^2\rf^2\ll1$ is assumed, we obtain
\begin{align}
\Aout&\simeq H\sin2\chis, \label{A-app}\\
\Cout&\simeq H\cos2\chis. \label{C-app}
\end{align}

Since the first radial null geodesic comes out from the star with $\chis=\chis|_{\etas=\etaf}\simeq\chii$, we have
\begin{align}
\Aout |_{\rm 1st}&\simeq H\sin2\chii, \label{A-1st}\\
\Cout |_{\rm 1st}&\simeq H\cos2\chii, \label{C-1st}
\end{align}
and
\begin{equation}
\kappa|_{\rm 1st}\simeq H\left(\frac{8\cos^2(2\chii)+4\cos(2\chii)+1}{4\cos(2\chii)+2}\right).
\end{equation}
For $0<\chii\ll1$, 
we have $\kappa|_{\rm 1st}\simeq13H/6$ and hence 
\begin{equation}
P|_{\rm 1st}\simeq \left(\frac{13}{6}\right)^2 P_{\rm GH},
\end{equation}
where $P_{\rm GH}$ is the radiation power of the Gibbons-Hawking radiation in the de Sitter spacetime with the Hubble constant $H$; 
\begin{equation}
P_{\rm GH}=\frac{H^2}{48\pi}.
\end{equation}

By using Eqs.~\eqref{chi-asympt} and \eqref{etaend-etac}, we have
\begin{equation}
\chis|_{\etas=\etaend}\simeq \frac{1}{2}\epsilon^{2\beta}.
\end{equation}
From this equation and Eqs.~\eqref{DB-ratio-last}, \eqref{A-app} and \eqref{C-app}, we obtain 
\begin{align}
\Aout|_{\rm last}&\simeq H\epsilon^{2\beta}, \\
\Cout&\simeq H
\end{align}
and hence 
\begin{equation}
\kappa|_{\rm last}\simeq H
\end{equation}
and
\begin{equation}
P|_{\rm last}=\frac{1}{48\pi}\kappa^2\simeq P_{\rm GH}.
\end{equation}
In the late stage of the radiative gravastar phase, $|\chis|$ becomes much less than unity. For $|\chis|\ll1$, $|\Aout|\ll H$ and $\Cout\simeq H$ hold  
and hence $\kappa\simeq H$. Thus, in this stage, the adiabatic condition $|\kappa'|\ll \kappa^2$ holds, which implies  
the thermal spectrum with the Gibbons-Hawking temperature
\begin{equation}
k_{\rm B}T_{\rm GH}=\frac{H}{2\pi}.
\end{equation} 

%%%%%%%%%%%%%%%%%
%%%%%%%%%%%%%%%%%
\subsection{Duration of radiative gravastar phase}
%%%%%%%%%%%%%%%%%
%%%%%%%%%%%%%%%%%

We estimate the duration of the radiative gravastar phase with respect to the time of the asymptotic observer. 
From Eqs.~\eqref{etaend-etac} and \eqref{HRG-approx}, we have
%\begin{align}
%&H\rf\sin^2(\etaend-\etac)-1-\cos(\etaend-\etac)\sqrt{1-H^2\rf^2\sin^2(\etaend-\etac)} \nonumber \\
%=&-\frac{(1-H^2\rf^2)\sin^2(\etaend-\etac)\sqrt{1-H^2\rf^2\sin^2(\etaend-\etac)}}
%{\sqrt{1-H^2\rf^2\sin^2(\etaend-\etac)}-\cos(\etaend-\etac)} \nonumber \\
%\simeq&-\frac{1}{8}\alpha\epsilon^{6\beta}\cot\chii,
%\end{align}
\begin{align}
&H\rf\sin^2(\etaend-\etac)-1-\cos(\etaend-\etac)\sqrt{1-H^2\rf^2\sin^2(\etaend-\etac)} \nonumber \\
=&\frac{(1-H\rf)^2\sin^2(\etaend-\etac)}
{H\rf\sin^2(\etaend-\etac)-1+\cos(\etaend-\etac)\sqrt{1-H^2\rf^2\sin^2(\etaend-\etac)}} \nonumber \\
\simeq&-\frac{1}{16}\alpha^2\epsilon^{8\beta}\cot^2\chii,
\end{align}
and
\begin{equation}
H\rf\sin^2(\etaend-\etac)+1-\cos(\etaend-\etac)\sqrt{1-H^2\rf^2\sin^2(\etaend-\etac)}\simeq 2.
\end{equation}
Thus, we have, from Eq.~\eqref{gravastar-t}, the time coordinate $t=\ts$ outside the star at $\eta=\etaend$ as 
%\begin{equation}
%\ts|_{\etas=\etaend}\simeq\frac{1}{2H}\ln\left(\frac{16\tan\chii}{\alpha\epsilon^{6\beta}}\right)+\tc.
%\end{equation}
\begin{equation}
\ts|_{\etas=\etaend}\simeq\frac{1}{2H}\ln\left(\frac{32\tan^2\chii}{\alpha^2\epsilon^{8\beta}}\right)+\tc.
\end{equation}
This is the time just before the 2nd burst occurs.  By contrast, since 
\begin{align}
\sin(\etaf-\etac)&=\sin\left(\pi-\chii+\frac{1}{2}\epsilon^2\right)=\sin\chii+{\cal O}(\epsilon^2), \\
\cos(\etaf-\etac)&=-\cos\chii+{\cal O}(\epsilon^2)
\end{align}
hold, we have
%\begin{align}
%&H\rf\sin^2(\etaf-\etac)-1-\cos(\etaf-\etac)\sqrt{1-H^2\rf^2\sin^2(\etaf-\etac)} \nonumber \\
%=&-\frac{(1-H^2\rf^2)\sin^2(\etaf-\etac)\sqrt{1-H^2\rf^2\sin^2(\etaf-\etac)}}
%{\sqrt{1-H^2\rf^2\sin^2(\etaf-\etac)}-\cos(\etaf-\etac)} \nonumber \\
%\simeq&-\frac{1}{2}\alpha\epsilon^{2\beta}\sin\chii\cos\chii,
%\end{align}
\begin{align}
&H\rf\sin^2(\etaf-\etac)-1-\cos(\etaf-\etac)\sqrt{1-H^2\rf^2\sin^2(\etaf-\etac)} \nonumber \\
=&\frac{(1-H\rf)^2\sin^2(\etaf-\etac)}
{H\rf\sin^2(\etaf-\etac)-1+\cos(\etaf-\etac)\sqrt{1-H^2\rf^2\sin^2(\etaf-\etac)}} \nonumber \\
\simeq&\frac{\alpha^2\epsilon^{4\beta}\cos^2\chii}
{4\left[H\rf\sin^2\chii-1-\cos\chii\sqrt{1-H^2\rf^2\sin^2\chii}\right]},
\end{align}
Thus, we get, from Eq.~\eqref{gravastar-t}, the time coordinate $t=\ts$ outside the star at $\eta=\etaf$ as 
%\begin{align}
%\ts|_{\etas=\etaf}
%&\simeq-\frac{1}{2H}\ln\left(\frac{1}{4}\alpha\epsilon^{2\beta}\sin\chii\cos\chii\right)+\tc.
%\end{align}
\begin{align}
\ts|_{\etas=\etaf}
&\simeq \frac{1}{2H}\ln
\left|\frac{4\left[(1+H^2\rf^2)\sin^2\chii-2-2\cos\chii\sqrt{1-H^2\rf^2\sin^2\chii}\right]}{\alpha^2\epsilon^{4\beta}\cos^2\chii}\right|
+\tc.
\end{align}
At this moment, the 1st burst ceases.  The duration $\Delta t$ between 1st and 2nd bursts, 
i.e., of the radiative gravastar phase  is then given as
\begin{equation}
\Delta t=\ts|_{\eta=\etaend}-\ts|_{\eta=\etaf}= \frac{1}{2H}\ln\epsilon^{-4\beta}+{\rm constant}
\simeq \frac{1}{H}\ln\epsilon^{-2\beta}. \label{duration-0}
\end{equation}

By definition, the value of $\epsilon$ determines when the slowing-down phase begins and ceases. 
Smaller the value of $\epsilon$ is, later the 1st burst occurs and the radiative gravastar phase starts. 
Furthermore, as shown in the above, the value of $\epsilon$ also determines the duration of the radiative gravastar phase.

From Eq.~\eqref{collapse-time}, the duration $\delta t$ with respect to the outside time coordinate $t$ 
from the beginning of the gravitational collapse, $\etas=0$, to a moment before the slowing-down phase,  
$0<\etas=\eta_{\rm bh}-2\varepsilon<\etab$, is given as 
\begin{align}
\delta t&=\frac{GM\cos\chii}{\sin^3\chii}\biggl[\left(1+2\sin^2\chii\right)(\pi-2\chii-2\varepsilon)+\sin(\pi-2\chii-2\varepsilon) \nonumber \\
&+\frac{2\sin^3\chii}{\cos\chii}\ln\left|\frac{\cot\chii+\cot(\chii+\varepsilon)}{\cot\chii-\cot(\chii+\varepsilon)}\right|\biggr]. 
\end{align}
Here we assume 
\begin{equation}
{\max}\left[\epsilon^{2\beta},~ \exp\left(-\frac{\pi}{2}\chii^{-3}\right)\right]\ll \varepsilon\ll\chii\ll1. \label{assmp}
\end{equation}
Then, we have 
\begin{equation}
\delta t \simeq \pi GM \chii^{-3}\simeq\sqrt{\frac{3\pi}{32G\rho|_{\etas=0}}}, \label{delta-t}
\end{equation}
where $\rho$ is the energy density of the dust. This result implies that 
the dust star shrinks to $\rs=2GM(1+2\chii^{-1}\varepsilon)\simeq2GM$ for the free-fall time. 

After the star collapsed to $\rs\simeq 2GM$,  the areal radius of the surface of the star is determined by 
\begin{equation}
\frac{d\rs}{d\ts}=-\frac{f(\rs)}{\cos\chii}\sqrt{\cos^2\chii-f(\rs)}\simeq-f(\rs)
\end{equation}
and thus, for $\ts>\delta t$, we have
\begin{equation}
\rs-2GM \simeq 4GM\chii^{-1}\varepsilon \exp\left(-\frac{\ts-\delta t}{2GM}\right).
\end{equation}
Denoting the time $\ts$ at the beginning of the slowing-down phase by $t_{\rm S}$, we have
\begin{equation}
\rb\simeq 2GM\left[1+2\chi^{-1}\varepsilon\exp\left(-\frac{t_{\rm S}-\delta t}{2GM}\right)\right].
\end{equation}
By comparing this equation with Eq.~\eqref{rb-app},  we have
\begin{equation}
\epsilon^{2\beta}\simeq 2\varepsilon\exp\left(-\frac{t_{\rm S}-\delta t}{2GM}\right).
\end{equation}  
Substituting this result into Eq.~\eqref{duration-0}, we obtain
\begin{equation}
\Delta t\simeq H^{-1}\left(\frac{t_{\rm S}-\delta t}{2GM}+\ln\varepsilon^{-1}\right).
\end{equation}
Here note that Eqs.~\eqref{assmp} and \eqref{delta-t} imply $\delta t\gg GM\ln\varepsilon^{-1}$. 
Thus, since $H^{-1}\simeq 2GM$ holds in the present model, 
if the period from the start of the gravitational collapse to the beginning time of the slowing-down phase is much longer 
than the free-fall time of the system, the duration of the radiative gravastar phase is nearly equal to $t_{\rm S}$.  

%%%%%%%%%%%%%%%%%
%%%%%%%%%%%%%%%%%
%%%%%%%%%%%%%%%%%
\section{Summary}
%%%%%%%%%%%%%%%%%
%%%%%%%%%%%%%%%%%
%%%%%%%%%%%%%%%%%

We studied the quantum particle creation in the gravastar formation process through the gravitational collapse 
of a spherically symmetric star. The star is assumed to be initially composed of homogeneously distributed dust 
and collapses in the freely falling manner. Just before the formation of the event horizon, the gravitational collapse stops 
due to the change of the equation of state of the star, and then the star eventually becomes the gravastar, in the present model. 

At the late stage of the gravitational collapse of the dust, the thermal radiation is generated by the quantum effect as 
pointed out by Hawking \cite{Hawking:1974,Hawking:1975}. The sudden stop of the gravitational collapse causes two bursts of the particle creation. 
The occurrence of two bursts was revealed by precedent researches in 
which a star with hollow inside and that occupied homogeneous matter 
enclosed by an infinitesimally thin crust were studied \cite{Harada-CM,Okabayashi-HN}. 
The characteristic behavior of the present gravastar formation model is 
that non-vanishing radiation is released in the period between 1st and 2nd bursts.  
Since this radiation is attributed to the gravastar, we have called this period the radiative gravastar phase. 
Its spectrum approaches the  thermal one with the Gibbons-Hawking temperature 
of the de Sitter spacetime inside the gravastar.  
Then, after the 2nd burst, no radiation is generated. 
The duration of the radiative gravastar phase is almost equal to $t_{\rm S}$, i.e., the period from the start of the gravitational collapse 
to the beginning of the slowing-down phase, if $t_{\rm S}$ is much longer than the free-fall time of the system. 
We should note that there is no black hole horizon and no cosmological horizon, but the transient thermal radiation appears.

\acknowledgments

KN is grateful to Hideki Ishihara and 
colleagues at the research groups of elementary particle physics and astrophysics in Osaka City University for 
useful discussions at colloquium. This work was supported by JSPS KAKENHI Grant
Number JP21K03557(KN), JP19K03876, JP19H01895, JP20H05853 (TH), and JP21J15676 (KO). 

\appendix
%%%%%%%%%%%%%%
%%%%%%%%%%%%%%
\section{Derivation of radiation power due to quantum effects}
%%%%%%%%%%%%%%
%%%%%%%%%%%%%%

In this section, we review the formulation given by Ford and Parker to derive the radiation power 
of quantum particle creation in general spherically symmetric gravitational collapse\cite{Ford-Parker}. 

The field operator $\hat{\phi}$ is represented as 
\begin{equation}
\hat{\phi}=\sum_{l,m}\int d\omega\left[a_{\omega lm}{\cal G}_{\omega lm}(t,\rstar,\Omega)
+a^\dagger_{\omega lm}{\cal G}^*_{\omega lm}(t,\rstar,\Omega)\right],
\end{equation}
where the mode function ${\cal G}_{\omega lm}$ satisfies the equation of motion for the scalar field 
\begin{equation}
\frac{1}{\sqrt{-g}}\partial_\mu\left(\sqrt{-g}g^{\mu\nu}\partial_\nu{\cal G}_{\omega lm}\right)=0,
\end{equation}
and the orthonormal conditions with respect to the Klein-Gordon norm
\begin{align}
({\cal G}_{\omega lm},{\cal G}_{\omega',l',m'})&:=-i\int \left[{\cal G}_{\omega lm}\left(\partial_\mu {\cal G}^*_{\omega',l',m'}\right)
+{\cal G}^*_{\omega',l',m'}\left(\partial_\mu {\cal G}_{\omega lm}\right)\right]\sqrt{-g}~n^\mu d\Sigma \nonumber \\
&=\delta(\omega-\omega')\delta_{ll'}\delta_{mm'},
\end{align}
where the integral is taken over a Cauchy surface and $n^\mu$ is the unit normal to this Cauchy surface. 
From the canonical commutation relation, we have
\begin{equation}
[a_{\omega lm},a_{\omega',l',m'}]=0,~~[a^\dagger_{\omega lm},a^\dagger_{\omega',l',m'}]=0,~~
[a_{\omega lm},a^\dagger_{\omega',l',m'}]=\delta(\omega-\omega')\delta_{ll'}\delta_{mm'}
\end{equation}
The vacuum state $|0\rangle$  is defined as the state which satisfies 
\begin{equation}
a_{\omega lm}|0\rangle=0. \label{vac-def}
\end{equation}

We adopt the mode functions which agree with those of the Minkowski spacetime in the past null infinity $u\rightarrow-\infty$. 
Hence, the mode functions in the asymptotic region $R\rightarrow\infty$ take the following form: 
\begin{equation}
{\cal G}_{\omega lm}=
\frac{1}{\sqrt{4\pi\omega}r}\left(e^{-i\omega F_{lm}(u)}+e^{-i\omega v}\right) Y_{lm}(\Omega), \label{asym-mode-function}
\end{equation}
where $Y_{lm}$ is the spherical harmonics. 
Here note that $F_{00}(u)$ is equal to $F(u)$ in Eq.~\eqref{sym-center}. Then, the state $|0\rangle$ defined 
as Eq.~\eqref{vac-def} is regarded as the Minkowski vacuum in the past null infinity. 

The stress-energy-momentum tensor operator of the scalar field is given as
\begin{equation}
\hat{T}_{\mu\nu}=(\partial_\mu\hat{\phi})\partial_\nu\hat{\phi}
-\frac{1}{2}g_{\mu\nu}g^{\alpha\beta}(\partial_\alpha\hat{\phi})\partial_\beta\hat{\phi}.
\end{equation}
The average energy flux is the expectation value of the following components of the stress-energy-momentum tensor: 
\begin{align}
\langle0|(\hat{T}_{uu}-\hat{T}_{vv})|0\rangle&=\langle0|\left[(\partial_u\hat{\phi})^2-(\partial_v\hat{\phi})^2\right]|0\rangle \nonumber \\
&=\sum_{lm}\int_0^\infty d\omega\left[\left(\partial_u {\cal G}_{\omega lm}\right)\partial_u {\cal G}^*_{\omega lm}
-\left(\partial_v {\cal G}_{\omega lm}\right)\partial_v {\cal G}^*_{\omega lm}\right].
\end{align} 
In order to obtain the average energy flux detected in the asymptotic region, we substitute Eq.~\eqref{asym-mode-function} into this equation, 
and replacing $u$ and $v$ by $u+\epsilon$ and $v+\epsilon$ in ${\cal G}^*_{\omega lm}$, 
we evaluate the integral with respect to $\omega$ as
\begin{align}
\langle0|(\hat{T}_{uu}-\hat{T}_{vv})|0\rangle 
&=\lim_{\epsilon\rightarrow0}\frac{1}{4\pi r^2}\sum_{lm}|Y_{lm}|^2\int_0^\infty d\omega \omega
\left[F_{lm}'(u)F_{lm}'(u+\epsilon)e^{i\omega\left[F_{lm}(u+\epsilon)-F_{lm}(u)\right]}-e^{i\omega\epsilon}\right] \nonumber \\
&\simeq\frac{1}{48\pi r^2}\left[\left(\frac{F''}{F'}\right)^2-2\left(\frac{F''}{F'}\right)'\right]|Y_{00}|^2,
\end{align}
where a prime represents a derivative with respect to $u$, and 
we have ignored the contributions from non-vanishing $l$ modes 
since those will not suffer the dynamical effect in the neighborhood of the center $r=0$ due to the centrifugal potential 
so much, and hence $|F''_{lm}|\ll |F_{00}''|=|F''|$ holds for $l>0$. Hereafter, we ignored the total derivative term $(F''/F')'$ in the last equality, 
which does not contribute the total emitted energy. 
The radiation power $P$ detected in the asymptotic region is obtained as
\begin{align}
P=\oint_{r\rightarrow\infty} \langle0|(\hat{T}_{uu}-\hat{T}_{vv})|0\rangle r^2d\Omega = \frac{1}{48\pi}\left(\frac{F''}{F'}\right)^2.
\end{align}

%%%%%%%%%%%%%%%%%
%%%%%%%%%%%%%%%%%
\section{Stress-energy-momentum of infinitesimally thin shell}
%%%%%%%%%%%%%%%%%
%%%%%%%%%%%%%%%%%

The Darmois-Israel junction condition specifies the stress-energy-momentum tensor 
of the infinitesimally thin shell on the surface of the star\cite{Israel:1966}.  

We adopt the orthonormal basis $\{\bm{e}_{(\alpha)}\}$ on the surface of the star; $\bm{e}_{(0)}$ is the unit tangent 
to the world line of an observer at rest on the stellar surface. 
Their components with respect to the coordinates just inside the star are given as
\begin{align}
e_{(0)}{}^\mu&=\left(\dot{\eta}_{\rm s},\dchis,0,0\right)=\left(\sqrt{\dchis^2+\frac{1}{\as^2}},\dchis,0,0\right) \\
e_{(1)}{}^\mu&=\left(\dchis, \sqrt{\dchis^2+\frac{1}{\as^2}},0,0\right), \\
e_{(2)}{}^\mu&=\left(0,0,\frac{1}{\as\sin\chis},0\right)=\left(0,0,\frac{1}{\rs},0\right), \\
e_{(3)}{}^\mu&=\left(0,0,0,\frac{1}{\as\sin\chis\sin\theta}\right)=\left(0,0,0,\frac{1}{\rs\sin\theta}\right),
\end{align}
whereas, with respect to the coordinates just outside the star, as
\begin{align}
e_{(0)}{}^{\mu'}&=\left(\dot{t}_{\rm s},\drs,0,0\right)=\left(\frac{\sqrt{\drs^2+f(\rs)}}{f(\rs)},\drs,0,0\right) \\
e_{(1)}{}^{\mu'}&=\left(\frac{\drs}{f(\rs)},\sqrt{\drs^2+f(\rs)},0,0\right), \\
e_{(2)}{}^{\mu'}&=\left(0,0,\frac{1}{\rs},0\right), \\
e_{(3)}{}^{\mu'}&=\left(0,0,0,\frac{1}{\rs\sin\theta}\right).
\end{align}
We define 
\begin{equation}
e^{(\alpha)\mu}=\eta^{(\alpha)(\beta)}e_{(\beta)}{}^\mu~~~~~{\rm and}~~~~~\eta_{(\alpha)(\beta)}e^{(\beta)\mu}=e_{(\alpha)}{}^\mu
\end{equation}
where $\eta^{(\alpha)(\beta)}={\rm diag}[-1,1,1,1]=\eta_{(\alpha)(\beta)}$. 

The components of the second fundamental form $K^{\rm (in)}_{\mu\nu}$ of the surface of the star with respect 
to the coordinates inside the star are given as
\begin{equation}
K^{\rm (in)}_{\mu\nu}=\left(e^{(0)}{}_\mu e_{(0)}{}^\alpha+e^{(2)}{}_\mu e_{(2)}{}_\alpha+e^{(3)}{}_\mu e_{(3)}{}^\alpha\right)\nabla_\alpha e_{(1)\nu}.
\end{equation}
By using this expression, we have the tetrad components of the second fundamental form in the form; 
\begin{align}
K^{\rm (in)}_{(0)(0)}&=-\frac{1}{\as\dot{\eta}_{\rm s}}\left(\ddot{\chi}_{\rm s}+\frac{2}{\as}\frac{d\as}{d\etas}\dot{\eta}_{\rm s}\dchis\right), \\
K^{\rm (in)}_{(2)(2)}&=\frac{1}{\as}\frac{d\as}{d\etas}\dchis+\cot\chis \dot{\eta}_{\rm s}=K^{\rm (in)}_{(3)(3)}
\end{align}
and all the other components vanish. 
By contrast, the components of the second fundamental form $K^{\rm (out)}_{\mu\nu}$ of the surface of the star with respect to 
the coordinates outside the star is given as
\begin{equation}
K^{\rm (out)}_{\mu'\nu'}=\left(e^{(0)}{}_{\mu'} e_{(0)}{}^{\alpha'}+e^{(2)}{}_{\mu'} e_{(2)}{}_{\alpha'}+e^{(3)}{}_{\mu'} e_{(3)}{}^{\alpha'}\right)\nabla_{\alpha'} e_{(1)\nu'}.
\end{equation}
Then, we obtain the tetrad components of the second fundamental form in the form
\begin{align}
K^{\rm (out)}_{(0)(0)}&=-\frac{1}{f(\rs)\dot{t}_{\rm s}}\left(\ddot{r}_{\rm s}+\frac{GM}{\rs^2}\right), \\
K^{\rm (out)}_{(2)(2)}&=\frac{f(\rs)}{\rs}\dot{t}_{\rm s}=\frac{1}{\rs}\sqrt{\drs^2+f(\rs)}=K^{\rm (out)}_{(3)(3)}
\end{align}
and all the other components vanish. 
By using Eqs.~\eqref{etas-dot}--\eqref{rs-ddot}, we concretely obtain $K_{(\alpha)(\beta)}^{\rm (in)}$ and $K_{(\alpha)(\beta)}^{\rm (out)}$. 

The Darmoise-Israel junction condition is
\begin{equation}
K^{\rm (out)}_{(\alpha)(\beta)}-K^{\rm (in)}_{(\alpha)(\beta)}=-8\pi G \left(S_{(\alpha)(\beta)}
-\frac{1}{2}h_{(\alpha)(\beta)}{\rm tr}S\right),
\end{equation}
where $h_{(\alpha)(\beta)}={\rm diag}[-1,0,1,1]$.  Then, the tetrad components, $S_{(\alpha)(\beta)}$, are given as
\begin{align}
S_{(0)(0)}&=-\frac{1}{4\pi G}\left(K^{\rm (out)}_{(2)(2)}-K^{\rm (in)}_{(2)(2)}\right), \\
S_{(2)(2)}&=-\frac{1}{8\pi G}\left(K^{\rm (out)}_{(0)(0)}-K^{\rm (out)}_{(2)(2)}-K^{\rm (in)}_{(0)(0)}+K^{\rm (in)}_{(2)(2)}\right)=S_{(3)(3)}
\end{align}
and the other components vanish. $S_{(0)(0)}$ is the energy per unit area, whereas $S_{(2)(2)}=S_{(3)(3)}$ is the tangential pressure.

%%%%%%%%%%%%%%
%%%%%%%%%%%%%%
\section{On the case of $\beta=1$}
%%%%%%%%%%%%%%
%%%%%%%%%%%%%%

We examine $|d\chis/d\etas|$ in the case of $\beta=1$. Here, we regard $\chis$ as a function of $\etas$ and the parameter $\alpha$. 
Then, from Eqs.~\eqref{cos-dchi-deta} to \eqref{1st-term}, we have 
\begin{equation}
 \frac{\partial\chis}{\partial\etas}\xrightarrow[\epsilon\rightarrow0]{}
-\gamma^3\left[70\alpha(1-\gamma)^3-70\gamma^3+216\gamma^2-225\gamma+80\right].
\end{equation}
Because of $0 < \gamma <1$, we have
\begin{equation}
\frac{\partial}{\partial \alpha} \left.\frac{\partial\chis}{\partial \etas}\right|_{\epsilon=0}=-70\gamma^3(1-\gamma)^3<0.
\end{equation}
Because of $0<\alpha<1$, the following inequality should hold: 
\begin{equation}
-\gamma^3\left[70(1-\gamma)^3-70\gamma^3+216\gamma^2-225\gamma+80\right]
<\frac{\partial\chis}{\partial\etas}<
-\gamma^3\left[-70\gamma^3+216\gamma^2-225\gamma+80\right].
\end{equation}

The rightmost side (RMS) of this inequality is rewritten in the form
\begin{equation}
{\rm RMS}=-\gamma^3\left[(1-\gamma)(70\gamma^2-146\gamma+79)+1\right].
\end{equation}
It is easy to see that $70\gamma^2-146\gamma+79>0$ holds for real $\gamma$. Hence we have RMS$<0$ and 
\begin{equation}
\left|\frac{\partial \chis}{\partial \etas}\right|_{\epsilon=0}>|{\rm RMS}|,
\end{equation}
for $0<\gamma<1$. It is easy to get
\begin{equation}
\frac{d}{d\gamma} |{\rm RMS}|=420\gamma^2(1-\gamma)^2\left(\frac{4}{7}-\gamma\right).
\end{equation}
This equation implies that $|$RMS$|$ has a maximum at $\gamma=4/7$ in $0<\gamma<1$, and the maximum value of $|$RMS$|$ is 
\begin{equation}
{\rm max}[|{\rm RMS}|]=\left|{\rm RMS}\right|_{\gamma=4/7}=\frac{27904}{16807}>1.
\end{equation}
This result implies that $|\partial\chis/\partial\etas|_{\epsilon=0}$ exceeds unity in the neighborhood of $\gamma=4/7$. 
Thus, the world line of the surface of the star cannot be kept timelike in the case of $\beta=1$ for any $0<\alpha<1$.

%%%%%%%%%%%%%%%%%%%%%%%%%%%%%
%%%%%%%%%%%%%%%%%%%%%%%%%%%%%
%%%%%%%%%%%%%%%%%%%%%%%%%%%%%
\section{Often used formulae}
%%%%%%%%%%%%%%%%%%%%%%%%%%%%%
%%%%%%%%%%%%%%%%%%%%%%%%%%%%%
%%%%%%%%%%%%%%%%%%%%%%%%%%%%%

Under the assumption of $0<\beta<1$, we often use the following approximations: 
\begin{align}
\rb&=\frac{GM}{\sin^2\chii}\left(1+\cos\etab\right)\simeq 2GM\left(1+\epsilon^{2\beta}\cot\chii\right), \label{rb-app}\\
\rf&=\frac{GM}{\sin^2\chii}\left[1+\cos\left(\etaf-\alpha\epsilon^{2\beta}\right)\right]\simeq 2GM\left(1+\alpha\epsilon^{2\beta}\cot\chii\right), \label{RG-approx}\\
H\rf&\simeq 1-\frac{1}{2}\alpha\epsilon^{2\beta}\cot\chii, \label{HRG-approx}\\
\chis|_{\etas=\etaf}&\simeq\chii-\frac{1}{2}\alpha\epsilon^{2\beta}. \label{chi-G}
\end{align}

%%%%%%%%%%%%%%
%%%%%%%%%%%%%%
\section{Equation of state}
%%%%%%%%%%%%%%
%%%%%%%%%%%%%%

\begin{figure}
\begin{center}
\includegraphics[width=0.6\textwidth]{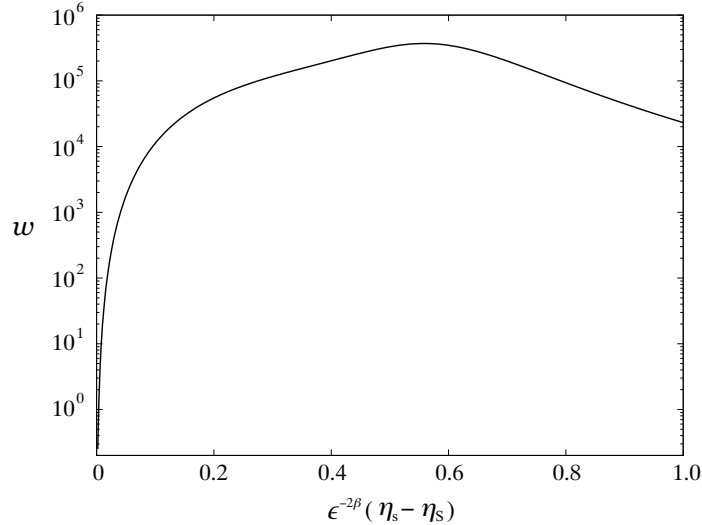}
\caption{\label{EOM}
The coefficient $w$ in the equation of state in the slowing-down phase of the example given in Sec.~IV 
is depicted as a function of normalized $\epsilon^{-2\beta}(\etas-\etab)$. It highly exceeds unity, and hence the 
dominant energy condition is not satisfied. }
\end{center}
\end{figure}

The equation of state of the star is nontrivial in the slowing-down phase. We write the equation of state 
in the form $p=w\rho$, where $p$ and $\rho$ are the pressure and the energy density, respectively. The Einstein equations imply
\begin{equation}
\frac{1}{a}\frac{d^2a}{d\tau^2}=-\frac{4\pi G}{3}(\rho+3p)~~~\Longrightarrow~~~
w=\frac{p}{\rho}=-\frac{1}{4\pi G\rho} \frac{1}{a}\frac{d^2a}{d\tau^2}-\frac{1}{3} 
\end{equation}
and 
\begin{equation}
\left(\frac{1}{a}\frac{da}{d\tau}\right)^2=\frac{8\pi G}{3}\rho-\frac{1}{a^2}~~~\Longrightarrow~~~
4\pi G\rho=\frac{3}{2a^2}\left[\left(\frac{da}{d\tau}\right)^2+1\right].
\end{equation}
Combining these equations, we have
\begin{align}
w&=-\frac{2a}{3}\left[\left(\frac{da}{d\tau}\right)^2+1\right]^{-1}
\frac{d^2a}{d\tau^2}-\frac{1}{3} \nonumber \\
&=-\frac{2}{3a}\left[\left(\frac{1}{a}\frac{da}{d\eta}\right)^2+1\right]^{-1}\left[\frac{d^2a}{d\eta^2}-\frac{1}{a}\left(\frac{da}{d\eta}\right)^2\right]-\frac{1}{3}.
\end{align}
 In order to see $w$ in the slowing-down phase, we may invoke Eq.~\eqref{da-deta} and 
 \begin{equation}
\frac{d^2\as}{d\etas^2}=\as\left[\frac{1}{\rs}\frac{d^2\rs}{d\etas^2}-\cot\chis\frac{d^2\chis}{d\etas^2}
-\frac{2}{\rs}\cot\chis\frac{d\rs}{d\etas}\frac{d\chis}{d\etas}
+\frac{1+\cos^2\chis}{\sin^2\chis}\left(\frac{d\chis}{d\etas}\right)^2\right]. \label{dda-ddeta}
\end{equation}
By Eqs.~\eqref{dchi-deta}--\eqref{ddrs-deta-lim}, we find 
\begin{equation}
\frac{d\as}{d\etas}=O(\epsilon^0)~~~~~{\rm and}~~~~\frac{d^2\as}{d\etas^2}=O(\epsilon^{-2\beta}),
\end{equation}
and hence
\begin{equation}
w=O(\epsilon^{-2\beta}).
\end{equation}
In Fig.~\ref{EOM}, we show the numerical result for $w$ in the slowing-down phase of the example given in Sec.~IV.  
The dominant energy condition will not be satisfied so that the gravitational collapse slows down just before the event horizon forms.

%%%%%%%%%%%%%
%%%%%%%%%%%%%
%%%%%%%%%%%%%
%%%%%%%%%%%%%

\end{document}